%%%%%%%%%%%%%%%%%%%%%%%%%%%%%%%%%%%%%%%%%%%%%%%%%%%%%%%%%%%%%%%%%%%%%%%%%%%%%
\input harvmac.tex
% Equations (overrides harvmac's equation macros)
\newcount\eqnum  
\eqnum=0
\def\eq{\eqno(\secsym\the\meqno)\global\advance\meqno by1}
\def\eqlabel#1{{\xdef#1{\secsym\the\meqno}}\eq }  

% References (overrides harvmac's reference macros)
\newwrite\refs 
\def\startreferences{
 \immediate\openout\refs=references
 \immediate\write\refs{\baselineskip=14pt \parindent=16pt \parskip=2pt}
}
\startreferences

\refno=0
\def\aref#1{\global\advance\refno by1
 \immediate\write\refs{\noexpand\item{\the\refno.}#1\hfil\par}}
\def\ref#1{\aref{#1}\the\refno}
\def\refname#1{\xdef#1{\the\refno}}

\newcount\exno
\exno=0
\def\Ex{\global\advance\exno by1{\noindent\sl Example \the\exno:

\nobreak\par\nobreak}}

\parskip=6pt

\overfullrule=0mm

%%%%%%%%%%%%%%%%%%%%%%%%%%%%
\def\frac#1#2{{#1 \over #2}}		

\let\ta=\theta
\let\d=\partial
\let\La=\Lambda

\def\Tt{{{\bf\tilde T}}}

\def\rw{{\rightarrow}}
\def\Tn#1{({\buildrel \leftarrow\over{\Tt^{#1}}})}

% References (overrides harvmac's reference macros)
\newwrite\refs
\def\startreferences{
 \immediate\openout\refs=references
 \immediate\write\refs{\baselineskip=14pt \parindent=16pt \parskip=2pt}
}
\startreferences

\refno=0
\def\aref#1{\global\advance\refno by1
 \immediate\write\refs{\noexpand\item{\the\refno.}#1\hfil\par}}
\def\ref#1{\aref{#1}\the\refno}
\def\refname#1{\xdef#1{\the\refno}}

%==========================================================================
% PAGE TITRE
\Title{\vbox{\baselineskip12pt
\hbox{LAVAL-PHY-{96-21}}}}
{\vbox {\centerline{The quantum SKdV$_{1,4}$ equation at $c=3$}
}}

\bigskip
\centerline{ P. Mathieu\foot{Work
supported by NSERC (Canada) } }

\smallskip\centerline{ \it D\'epartement de
Physique, Universit\'e Laval, Qu\'ebec, Canada G1K 7P4}
\vskip .2in
\bigskip
\bigskip
\bigskip
\bigskip
\bigskip
\centerline{\bf Abstract}
\bigskip
\noindent
At $c=3$, two of the three integrable quantum $N=2$ supersymmetric Korteweg-de
Vries equations become identical (SKdV$_1$ and SKdV$_4$).  Quite remarkably, all
their conservation laws can be written in  closed form, which provides thus a
simple constructive integrability proof.

\Date{06-96}
%========================================================================

The quantum extension of the usual Korteweg-de Vries (KdV) equation is rather
easily formulated [\ref{B.A. Kuperschmidt and P. Mathieu, Phys. Lett. {\bf B227}
(1989) 245.}, \ref{R. Sasaki and I. Yamanaka, Adv. Stud. in Pure Math. {\bf 16}
(1988) 271.}\refname\sasa ]: it is the canonical equation obtained from the
quantum version of the KdV second Hamiltonian structure 
(i.e., the energy-momentum
tensor OPE) and the normal ordered form of the corresponding KdV Hamiltonian:
$${\dot T} = -[H,T], \qquad H={1\over 2\pi i}\oint dz~(TT)$$
Since the classical Hamiltonian has a single term, its quantum form is not
ambiguous.  The quantum extension of the supersymmetric KdV equation is also 
defined unambiguously [\ref{P. Mathieu, Nucl. Phys. {\bf B336} (1990) 338.},
\ref{I. Yamanaka and R. Sasaki, Prog. Theor. Phys. {\bf 79} (1988) 1167.}].
  This
is no longer true for the three integrable  $N=2$ supersymmetric KdV (SKdV)
equations  [\ref{P. Mathieu and M.A. Walton, Phys. Lett. {\bf B254} (1991)
106.}\refname\MW] since the three different Hamiltonian contains two terms, 
hence
a troublesome relative coefficient whose quantum form cannot be fixed a priori.
More explicitly, this  classical Hamiltonian is [\ref{C.-A. Laberge and 
P. Mathieu,
Phys. Lett. {\bf B215} (1988) 718; P. Labelle and P. Mathieu, J. Math. Phys.
{\bf 32} (1991) 923.}\refname\LL]:
$$H_{\alpha}^{\rm class} = \int dX\;\left \{{\cal U}^3
-{3\over 2\alpha}\;({\cal U}[D^+,D^-]{\cal U})\right\}$$
where ${\cal U}(x,\ta^+,\ta^-,t)$ is the classical form of the $N=2$
superenergy-momentum tensor, $dX=dxd\ta^-\ta^+$ and 
$D^\pm = \d_{\ta^\mp}+\ta^\pm\d_x$.  The corresponding KdV equation is the
canonical equation obtained from this Hamiltonian and  the Poisson bracket
version of the
$N=2$ superconformal algebra. It is integrable for only three values of
$\alpha$, namely $1,-2, 4$ [\LL, \ref{Z. Popowicz, Phys. Lett. {\bf A174} 
(1994) 411. }]. 

An elegant way of determining the quantum form of these integrable equations is
based on their correspondence with the (conjectured) integrable
perturbations of the
$N=2$ minimal models with $c=3K/( K+2)$ [\MW, \ref{P. Fendley, S.D. Mathur, C.
Vafa and N. Warner, Phys. Lett. {\bf B243} (1990) 257.}, \ref{P. Fendley, W.
Lerche, S.D. Mathur and N. P. Warner, Nucl. Phys. {\bf B348} (1991)
66.}\refname\Fen].
 This fixes the relative  
coefficient of the model defining quantum Hamiltonian.  
Identifying the equations
by the corresponding perturbation (that is, the chiral field label $\ell$ which
represents the perturbation
$\Phi^\ell$) as well as with the corresponding value of the classical label
$\alpha$, we have [\MW]\foot{Further
support for the conjectured integrability of the $\ell=1$ and $\ell= K$
perturbations is presented in [\ref{P. Di Francesco and P. Mathieu, Phys.
Lett.{\bf B278} (1992) 79.}]  and in [\ref{T. Eguchi and S.K. Yang, Mod. Phys.
Lett. {\bf A5} (1990) 1693.}] respectively.  Moreover, for 
each value of $K$, the
perturbations
$\ell=1,2$ have been related to quantum affine Toda theories at 
a particular value
of the coupling [\Fen] and these Toda models have been proved 
to be integrable in
[\ref{B. Feigin and E. Frenkel, {\it Integrals of motion and quantum groups},
 proceedings of C.I.M.E. Summer School on `Integrable systems
and Quantum groups', 1993
{ hep-th/9310022}.}]. }
$$\eqalign{
& \ell =1\;\,\qquad  H_{\alpha=4} = \oint dZ \;\left\{(\Tt(\Tt\Tt))
+\frac{1}{16}\;(c-3)\;(\Tt[D^+,D^-]\Tt)\right\}\cr
& \ell =2\;\,\qquad  H_{\alpha=1} = \oint dZ \;\left\{(\Tt(\Tt\Tt))
+\frac{1}{4}\;(c-3)\;(\Tt[D^+,D^-]\Tt)\right\}\cr
& \ell =K\qquad  H_{\alpha=-2} = \oint dZ \;\left\{(\Tt(\Tt\Tt))
-\frac{1}{8}\;(c-12)\;(\Tt[D^+,D^-]\Tt)\right \}\cr}$$
where $\Tt(Z)$ is the $N=2$ superenergy-momentum tensor 
$$\Tt(Z) = J(z)+\frac12 \ta^- G^+(z)-\frac12\ta^+ G^-(z) + \ta^+\ta^- T(z)$$
whose OPE reads
$$\Tt(Z_1)\;\Tt(Z_2) = {c/12\over Z_{12}^2}+{\ta_{12}^+
\ta_{12}^-\; \Tt(Z_2)\over
Z_{12}^2} + {\ta_{12}^+\;D^-\Tt(Z_2)\over 2 Z_{12}} - 
{\ta_{12}^-\;D^+\Tt(Z_2)\over 2
Z_{12}} +{\ta_{12}^+\ta_{12}^-\; \d\Tt(Z_2)\over Z_{12}}$$
with $Z_{12}\equiv z_1- z_2-\ta_1^+\ta_2^--\ta_1^-\ta_2^+$.

  In the classical limit ($c\rw\pm\infty$), the
ratios of the three relative coefficients are seen to be the same as the
inverse ratios of the quoted values of $\alpha$ and  the classical Hamiltonians
are recovered by the relation $\Tt = -c\,{\cal U}/6$.

Denote by qSKdV$_{\alpha}$ the equations obtained canonically from the
quantum Hamiltonian
$H_{\alpha}$.  Since at $c=3$, $H_{\alpha=4}= H_{\alpha=1}$
\foot{By treating this
value of $c$ as the limiting minimal model with $K\rw \infty$, 
we further verify
that the degenerate equation of the perturbing field $\Phi^2$ 
is a descendant of
that of  $\Phi$ (both perturbations have vanishing conformal dimension at
$c=3$).}, the two equations qSKdV$_{1}$ and qSKdV$_{4}$ merge into a single
one.\foot{Similarly, when $c=6$,  
$H_{\alpha=1}= H_{\alpha=-2}$ and for $c=9$, $H_{\alpha=4}=H_{\alpha=-2}$. 
Moreover, when $c=1$ or $3/2$ , thanks to the vacuum singular vector:
$$\eqalign{
&  c=1: \quad \left(J_{-1}^2-\frac16 L_{-2}\right)\;|0\rangle\cr
&  c=\frac32: \quad \left(J_{-1}^3-\frac38 J_{-1}L_{-2} -\frac1{16}
J_{-3}  -\frac{3}{64}L_{-3} + \frac{3}{64}G^+_{-3/2}G^-_{-3/2}
\right)\;|0\rangle\cr}$$ the three Hamiltonians reduce to $\oint dZ\;
(\Tt(\Tt\Tt))$.
}  
It turns out
that their conserved densities have an extremely simple form, namely
$$ \Tn{n} = (\cdots(((\Tt \Tt)\Tt)\Tt)\cdots \Tt)\qquad
(n~\hbox{factors})\eqlabel\cl$$ Hence, every conserved density 
has a single term
but normally ordered toward the left. This is the exact analog of the qKdV
conservation laws at $c=-2$ [\sasa, \ref{P. Di Francesco, P. Mathieu and D.
S\'en\'echal, Mod. Phys. Lett. {\bf A7} (1992) 701.}]. The rest of this note is
devoted to the proof of this result, which boils down to a simple exercise in
normal ordering rearrangements. 

The idea of the proof is to use the $c=3$ free field representation [\ref{G.
Mussardo, G. Sotkov and M. Stanishkov, Int. J. Mod. Phys. {\bf A4} (1989) 1135.}] :
$$\Tt = -\frac14(D^+S_+\,D^-S_-)$$
As usual parentheses denote normal
ordering.  The free field OPE's are
$$S_+(Z_1)\,S_-(Z_2) \sim -\ln Z_{12}+{\ta_{12}^+\ta_{12}^-\over Z_{12}}$$
and $S_+ S_+\sim S_-S_-\sim0$. $S_+$ and $S_-$ are chiral primary fields: 
$D^-S_+=D^+S_-=0$. The canonical equations of these fields take an extremely
simple form. As a result, the explicit expression for all 
the conserved charges of
this chiral free field system can be written down readily. 
The final step amounts
to prove that these can be reexpressed in terms of $\Tt$, according to (\cl). 

 At $c=3$, the Hamiltonian for the model under consideration is
$${\tilde H}_3 =\oint dZ\; (\Tt(\Tt\Tt)) = -\frac3{8}\oint
dZ\; (S_+^{(3)}\,S_-)$$
where  $S_+^{(n)} = \d^n S_+$. The canonical equations for the fields $S_+,
S_-$
$${\dot S}_+ = -[{\tilde H}, S_+]\;, \qquad {\dot S}_- = -[{\tilde H}, S_-]$$
 reduce to (with an appropriate time rescaling):
$${\dot S}_+ = S_+^{(3)}\;, \qquad {\dot S}_- = S_-^{(3)}$$
The infinite sequence of conserved charge for this system reads then
$${\tilde H}_n = \oint
dZ\; (S_+^{(n)}\,S_-)$$
where $n$ is any positive integer. It is simple to 
verify that $[{\tilde H}_n ,
{\tilde H}_m ]=0$. Now, as for the KdV equation at
$c=-2$, these charges can be written in terms of the normal ordered powers
of $\Tt$, but with a left nesting.  This is the announced result:
$${\tilde H}_n = \oint  dZ\; \Tn{n}$$
which  will be proved by recursion.

Normal ordering of $N=2$ superfield is defined as
$$({\cal A} {\cal B})(Z_2)= {1\over 2\pi i}\oint dZ_1\;
{\ta_{12}^+\ta_{12}^-\over
Z_{12}} {\cal A}(Z_1)\, {\cal B}(Z_2)$$ Note in particular that if 
$${\cal A}(Z_1)\; {\cal B}(Z_2) = \sum_{n=N}^{-\infty} Z_{12}^{-n} \left[{\cal
C}_n(Z_2)
 + \ta_{12}^+{\cal D}_n(Z_2) +
\ta_{12}^-{\cal E}_n(Z_2) +\ta_{12}^+\ta_{12}^-{\cal F}_n(Z_2)\right]$$
the  normal ordered commutator (anticommutator 
if $ {\cal A}$ and $ {\cal B}$ are 
both fermionic) is
$$([{\cal A}, {\cal B}])(Z_2) =\sum_{n>0}{(-1)^n\over n!} 
\d^n {\cal C} _n(Z_2)$$ 
Standard reordering manipulations rely on the rearrangement lemmas
of [\ref{F.A. Bais, P. Bouwknegt, K. Schoutens and M. Surridge, Nucl.
Phys. {\bf B304} (1988) 348.}], e.g., 
$$((AB)(CD)) = (C(D(AB)))+ (([(AB),C ])D) + (C([(AB),D]))$$
To simplify somewhat the notation, we will set 
$$\Lambda_\pm \equiv D^\pm  S_\pm$$ 
We easily
find that
$$\eqalign {
& \Tn{2} = (\Tt \Tt) =
-\frac18\left[(\La_+^{(1)}\La_-) - (\La_+\La_-^{(1)})\right]\cr & 
\Tn{3} = ((\Tt \Tt)\Tt) =
-\frac3{32}\left[(\La_+^{(2)}\La_-)+(\La_+\La_-^{(2)})\right] \cr}$$
Let us then assume that $\Tn{n}$ has the form
$$\Tn{n} = b_n\; \left[( \La_+^{(n-1)}\,\La_-)+(-1)^{n-1} (\La_+
\,\La_-^{(n-1)})\right]\eqlabel\tt$$
We now prove that
$$\Tn{n+1} = (\Tn{n}\Tt) =\left ({n+1\over 2n}\right) b_n \left[(
\La_+^{(n)}\,\La_-)+(-1)^{n} (\La_+
\,\La_-^{(n)})\right]\eqlabel\tnn$$ 
The proof can  be worked out in few lines. 
$\Tn{n+1}$ is equal to  
$$\Tn{n+1} = 
 -{b_n\over 4}[\,\Gamma_1+(-1)^{n-1}\Gamma_2\,]\eqlabel\sag$$ where
$$\eqalign{
  \Gamma_1=~& \left((\La_+^{(n-1)}\,\La_-)\,( \La_+ \La_-)\right)\cr
 = ~&
 \left(([\,\La_+^{(n-1)}\La_-,\, \La_+])\,\La_-\right)+ 
\left(\La_+\,([\,\La_+^{(n-1)}\La_-,\,\La_-])\right) \cr
=~&  
-2\,( \La_+^{(n)}\La_-) +(-1)^{n+1} {2\over n}\,(\La_+\La_-^{(n)})\cr
\Gamma_2=~&\left((\La_+\,\La_-^{(n-1)})\,( \La_+ \La_-)\right)\cr 
=~& 
\left(([\,\La_+\La_-^{(n-1)},\,\La_+])\,\La_-\right)+\left(\La_+
\,([\,\La_+\La_-^{(n-1)},\,
\La_-])\right) \cr  
=~&  (-1)^{n}{2\over
n}\,(\La_+^{(n)}\La_-)+2\,( \La_+\La_-^{(n)})\cr}\eqlabel\det$$
The substitution
of (\det) into (\sag)  yields (\tnn). The recursion argument 
proves (\tt) and fixes the coeffficient $b_n$:
$$b_n = -{n \over 2^{n+2}}$$

Notice that the quartic terms disappear in $\Tn{n+1}$ thanks to the
fermionic character of the
$\La_\pm$. For this it is  crucial that $\Tn{n}$ be composed of terms
which all contain a factor $\La_+$ or $\La_-$ without derivatives. This
property is lost when the powers of $\Tt$ are ordered toward the right (and as
a result, the conserved densities no longer have a simple form).

Up to a total derivative, $\Tn{n}$ is 
proportional to $(S_+^{(n)}\,S_-)$.  As already indicated, 
the integrals ${\tilde
H}_n $ are mutually commuting. We have thus obtain a  
rigorous and constructive 
integrability proof for the qSKdV$_{1,4}$ equation at $c=3$.

For the other SKdV equation, the presence of the term 
$(\Tt[D^+, D^-]\Tt)$ in the
Hamiltonian induces quartic contribution in the fields 
$S_+, S_-$ that generates
cubic terms in the equations for $S_+$ and $S_-$.  
These terms couple the two
fields and the equation become sufficienty complicated to prevent a closed form
expression for their conservation laws.

\immediate\closeout\refs \vskip 0.5cm
  \message{References}\input references
\end